\documentclass[12pt]{article}
 
\begin{document}

\renewcommand{\theequation}{\arabic{section}.\arabic{equation}}
\baselineskip=20pt


\title{\bf su(2) and su(1,1) displaced number states
       and their nonclassical properties}

\author{                   
       Hongchen Fu$^{\dagger\S}\thanks{E-mail: h.fu@open.ac.uk}$ \ 
       Xiaoguang Wang$^{\ddagger}$, Chong Li$^{\S}$ and
       Jiangong Wang$^{\S}$ \\ \  \\
{\normalsize \it                       
       $\dagger$   Quantum Processes Group, The Open University,      
       Milton Keynes, MK7 6AA, U.\,K. }\\  
{\normalsize \it         
       $\ddagger$ Institute of Physics and Astronomy,
       University of Aarhus, DK-8000 Aarhus, Denmark }\\  
{\normalsize \it         
       $\S$  Institute of Theoretical Physics, Northeast
       Normal University,}\\
{\normalsize \it           
       Changchun 130024, P.\,R.\,China  }} 

\maketitle

\begin{abstract}
We study su(2) and su(1,1) displaced number states. Those 
states are eigenstates of density-dependent interaction systems
of quantized radiation field with classical current. Those states
are intermediate states interpolating between number and 
displaced number states. Their photon number distribution, 
statistical and squeezing properties are studied in detail. It is show
that these states exhibit strong nonclassical properties.
\end{abstract}

\section{Introduction}

Coherent states and their various generalizations play
very important roles in many fields of physics \cite{CS}.
They were first discovered in 1926 by Schr\"{o}dinger, as
an example of non-spreading wave packet \cite{sch}. In
1950's, Senitzky, Plebanski, Husimi and Epstein found other
wave packets \cite{Sen} , which keep their shape and follow 
the classical motion, before the modern work of
coherent states by Glauber and Sudarshan in 1963 \cite{Gla}. 
These wave packets are essentially the {\it displaced number 
states}, $ D(\alpha)|n\rangle$, where 
$ D(\alpha)\equiv \exp(\alpha a^{\dagger} -\alpha^* a)$ 
is the displacement operator and $a^{\dagger} $, $ a $
are the creation and annihilation operators
satisfying $[a, a^{\dagger}]=1$.

Since Stoler {\it et al.} introduced the binomial states (BS)
 in 1985 \cite{BS1}, the so-called {\it intermediate states} which 
interpolate between two fundamental quantum optical states
have attracted much attention \cite{inters}. In a previous
paper \cite{GBS} 
Sasaki and one of authors of this paper proposed the 
the {\it intermediate number-squeezed states}. 
One of these intermediate states is the so-called
su(2) {\em displaced number states} (DNS) 
\begin{equation}
     |M,\xi,n\rangle\equiv D_2(M,\xi) |n\rangle,  \qquad 
     D_2(M,\xi)\equiv \exp(\xi J_M^+-\xi^* J_M^-),
\end{equation}     
in which $|n\rangle$ ($0\leq n \leq M$) is the number state and 
\begin{equation}
    J_M^+=a^{\dagger}\sqrt{M-N},\quad
    J_M^-=\sqrt{M-N}a, \quad
    J_M^0=N-M/2
\end{equation}
are the Holstein-Primakoff (HP) realization of su(2) 
algebra. It is obvious that $|M,\xi,n\rangle$ is a finite 
superposition of number states $|n\rangle$ for 
$n=0,1,\cdots,M$.

One can also define the {\em non-compact} counterpart of 
the su(2) DNS, the {\em su(1,1) displaced number state}
\begin{equation}
    \|M, \zeta, n\rangle \equiv D_{11}(M,\zeta)|n\rangle,\quad
    D_{11}(M,\zeta)\equiv \exp(\zeta K_M^+ -\zeta^* K_M^-),
    \label{dns11}
\end{equation}
where $K_M^+, K_M^-$ and $K_M^0$ are generators of su(1,1) Lie algebra
via its HP realization
\begin{equation}
    K_M^+=a^\dagger \sqrt{M+N},\ \ \ \ \
    K_M^-=\sqrt{M+N} a,\ \ \ \ \
    K_M^0=N+ M/2,
\end{equation}
and $M/2$ ($M$ is non-negative integer) is the Bargmann index. 
This is a natural generalization of the negative binomial
states \cite{nbs} which correspond to $ n=0 $.

In this paper we shall prove that both su(2) and 
su(1,1) DNS are eigenstates of 
density-dependent interaction systems of quantized 
radiation field with classical current, and 
that both are {\it intermediate 
states} between the number and the displaced number 
states and reduce to them in two different limits. 
We also study their expansion in Fock space and 
their photon number distributions, statistical properties
and their squeezing effect. It is of interest 
that these states which interpolate between 
number and displaced number states, neither of 
which exhibit squeezing, are nevertheless squeezing states.
 
\section{Physical interpretation}
\setcounter{equation}{0}

\subsection{Physical systems}

In this section we shall see that both su(2) and su(1,1) DNS
 are eigenstates of some interaction systems of
quantum radiation field with a classical current. To see this,
we first see that the su(2) and su(1,1) DNS
satisfy the following eigenvalue equations
(writing $\xi\equiv re^{i\theta}, \zeta=Re^{i\vartheta}$)
\begin{equation}
    H_2 |M,\xi,n\rangle = E_n(2)|M,\xi,n\rangle,  \quad
    H_{11} \|M,\zeta,n\rangle = E_n(1,1)\|M,\zeta,n\rangle,
\end{equation}
where $H_2$ and $H_{11}$ 
\begin{eqnarray}
    && H_{2}=\omega N -\frac{\omega}{2}\tan(2r)\left(
    e^{i\theta} a^\dagger \sqrt{M-N} +
    e^{-i\theta}\sqrt{M-N}a\right),\\ &&
    H_{11}=\omega N -\frac{\omega}{2}\tanh(2R)\left(
    e^{i\vartheta} a^\dagger \sqrt{M+N} +
    e^{-i\vartheta}\sqrt{M+N}a\right),
\end{eqnarray}
are Hermitian operators and can be cased as 
the Hamiltonians describing
interaction between a single mode radiation field
with frequency $\omega $ and  a classical current and this
interaction is density-dependent, and $E_n(2)$
and $E_n(11)$ are eigenvalues (energy)
\begin{equation}
    E_n(2)=\frac{M}{2}+\frac{2n-M}{2\cos(2r)},\quad
    E_n(11)= \frac{M+2n}{2\cosh(2R)}-\frac{M}{2}.
\end{equation}   
Requirement of non-negative energy gives
\begin{eqnarray}
   &&\mbox{For su(2) case: } \cos(2r)>\max(0, 1-2n/M)\mbox{ or }
                  \cos(2r)<\min(0, 1-2n/M); \nonumber \\
   &&\mbox{For su(1,1) case: } \cosh(2R)\leq 1+\frac{2n}{M}.
        \nonumber
\end{eqnarray}

\subsection{Limiting states}

We now prove that both su(2) and su(1,1) DNS
are intermediate states interpolating between number
and displaced number states of radiation field. For su(2) case,
it is easy to see that, in the limit 
$ M\to\infty $, $|\xi|\to 0 $
with $|\xi|^2 M=\alpha^2 $ ($\alpha$ real) 
fixed and $n$ finite 
\begin{equation}
    \xi J_M^+ \to \alpha e^{i\theta} a^{\dagger},\quad
    \xi^* J_M^-\to \alpha e^{-i\theta} a, \quad
    D_2(M,\xi)\to D(\alpha e^{i\theta})\equiv
    e^{\alpha (e^{i\theta}a^\dagger- e^{-i\theta}a)},
\end{equation}
and thus the su(2) DNS degenerate to the
displaced number state of radiation field
\begin{equation}
    |M, \xi, n\rangle\longrightarrow D(\alpha e^{i\theta})
    |n\rangle =
    e^{\alpha (e^{i\theta}a^\dagger- e^{-i\theta}a)}
    |n\rangle,
\end{equation}
while in a different limit $\xi\to m\pi e^{i\theta}$ with $m$ integer, 
we have $D_2(m\pi e^{i\theta})\to 1$ and 
the su(2) DNS degenerates to the 
number state $ |n\rangle $ of radiation field.

Similarly, the su(1,1) DNS tend to the
displaced number states of radiation field in the limit
$ M\to\infty $, $|\zeta|\to 0$ keeping $M|\zeta|^2=\alpha^2$, and 
to the number state $|n\rangle $ in the  limit $\zeta\to
m\pi e^{i\vartheta}$.

\section{Photon number distributions}
\setcounter{equation}{0}

\subsection{su(2) case}

The su(2) DNS can be
expanded as a finite linear combination of Fock states
\begin{equation}
    D_2(M,\xi)|n\rangle =\sum_{m=0}^{M}\langle m|D_2(M,\xi)|n\rangle
    |m\rangle
\end{equation}
and what we need to do is to determine the matrix element
$\langle m|D_2(M,\xi)|n\rangle $. Using the disentangling theorem
of su(2)
\begin{equation}
    D_2(M,\xi)=\exp(\xi J_M^+ -\xi^* J_M^-)=\exp(\varsigma J_M^+)
    (1+|\varsigma|^2)^{J_M^0}
    \exp(-\varsigma^* J_M^-),
\end{equation}
where $ \varsigma=e^{i\theta}\tan r $, the matrix elements 
can be written as
\begin{equation}
    \langle m|D_2(M,\xi)|n\rangle=\langle m|
    \exp(\varsigma J_M^+)(1+|\varsigma|^2)^{J_M^0}
    \exp(-\varsigma^* J_M^-)|n\rangle. \label{matrix111}
\end{equation}
From
\begin{equation}
    (J_M^-)^k|n\rangle=\left\{ \begin{array}{ll}
             \sqrt{\frac{\displaystyle (M-n+k)!n!}{\displaystyle
             (M-n)!(n-k)!}} |n-k\rangle,  & k\leq n,\\
             0,  & k>n,
             \end{array}\right.
\end{equation}
we have
\begin{eqnarray}
    \exp(-\varsigma^* J_M^-)|n\rangle &=&\sqrt{\frac{n!}{(M-n)!}}
        \sum_{k=0}^n \frac{(-\varsigma^*)^{(n-k)}}{(n-k)!}
        \sqrt{\frac{(M-k)!}{k!}}|k\rangle , \label{111}\\
    \langle m|\exp(\varsigma J_M^+) &=&\sqrt{\frac{m!}{(M-m)!}}
        \sum_{k=0}^m \frac{(\varsigma^*)^{(m-k)}}{(m-k)!}
        \sqrt{\frac{(M-k)!}{k!}}\langle k|.\label{222}
\end{eqnarray}
Inserting Eqs.\,(\ref{111}, \ref{222}) into Eq.\,(\ref{matrix111})
we finally obtain the desired matrix elements
\begin{equation}
   \langle m|D_2(M,\xi)|n\rangle=e^{i\theta(m-n)}
   {\cal D}_{m}(M,n,|\varsigma|), 
    \label{matrixelement}
\end{equation}  
where
\begin{equation}
    {\cal D}_{m}(M,n,|\varsigma|)\equiv(-1)^n |\varsigma|^{m+n}\left[
    \frac{m!n!}{(M-m)!(M-n)!}\right]^{1/2}\,
    F(M,n,m,|\varsigma|^2),
\end{equation}
and 
\begin{equation}
    F(M,n,m,|\varsigma|^2)=\sum_{k=0}^{\min(m,n)} 
    \frac{(M-k)!(-|\varsigma|^2)^{-k}(1+|\varsigma|^2)^{k-M/2}}{
    k!(n-k)!(m-k)!}
\end{equation}
is a Hypergeometric function. 
  
The photon distribution is easily obtained as
\begin{equation}
   P_m(M,n,|\varsigma|^2)=\frac{m!n!}{(M-m)!(M-n)!}|\varsigma|^{2(M+n)}
   \left[F(M,n,m,|\varsigma|^2)\right]^2.
\end{equation}
In the special case $n=0$, $P_m(M,n,|\varsigma|^2)$ reduces to the
binomial distribution \cite{bsd} 
\begin{equation}
   P_m(M,0,|\varsigma|^2)={M\choose n}p^n (1-p)^{M-n},\ \ \ \ \ 
   p=\frac{|\varsigma|^2}{1+|\varsigma|^2},
\end{equation}
as we expected.

\subsection{su(1,1) case}

Using the disentangling theorem of su(1,1) 
(writing $\lambda=e^{i\vartheta}\tanh R$)
\begin{equation}
    D_{11}(M,\zeta)=\exp(\zeta K_+ -\zeta^* K_-)=
    \exp(\lambda K_+)(1-|\lambda|^2)^{K_0}\exp(-\lambda^* K_-),
\end{equation}
and
\begin{eqnarray}
    \exp(-\lambda^* K_-)|n\rangle &=&\sqrt{n!(M+n-1)!}
    \sum_{k=0}^n \frac{(-\lambda^*)^{n-k}}{(n-k)!
    \sqrt{k!(M+k-1)!}}|k\rangle, \nonumber \\
    \langle m|\exp(\lambda K_+)&=&\sqrt{m!(M+m-1)!}
    \sum_{k=0}^m \frac{\lambda^{m-k}}{(m-k)!
    \sqrt{k!(M+k-1)!}}\langle k|,
\end{eqnarray} 
we obtain 
\begin{equation}
    \langle m| D_{11}(M,\zeta)|n\rangle = e^{i\vartheta (m-n)}
    {\cal G}_m(M,n,|\lambda|),
\end{equation}
where
\begin{equation}
    {\cal G}_m(M,n,|\lambda|)\equiv(-1)^n|\lambda|^{m+n}
    \sqrt{m!n!(M+m-1)!(M+n-1)!}G(M,n,m,|\lambda|^2),
\end{equation}
and
\begin{equation}
    G(M,n,m,|\lambda|^2)=\sum_{k=0}^{\min(m,n)}
    \frac{(-|\lambda|^2)^{-k}(1-|\lambda|^2)^{k+M/2}}{
    k!(M+k-1)!(n-k)!(m-k!)},
\end{equation}
is also a hypergeometric function.

The photon distribution is then obtained as
\begin{equation}
    P_{11}(m)=
    m!n!(M+m-1)!(M+n-1)! |\lambda|^{2(M+n)}G^2(M,n,m,|\lambda|^2).
\end{equation}
When $n=0$, $ P_{11}(m)$ reduces to the
negative binomial distribution as we expected
\begin{equation}
    P_{11}(m)={M+m-1 \choose m}|\lambda|^{2m}(1-|\lambda|^2)^M.
\end{equation}

\section{Photon statistics}
\setcounter{equation}{0}

\subsection{su(2) case}

Writing
$  \langle M,\xi,n |N|M,\xi,n\rangle=
    \langle n|D^{-1}_2(M,\xi)
    J_M^0 D_2(M,\xi)|n\rangle +M/2, 
$
where $N\equiv a^\dagger a$,  and using
\begin{equation}
D^{-1}_{2}(M,\xi)J_M^0 D_2(M,\xi)=\frac{1}{2}
       (e^{i\theta}J_M^+ + e^{-i\theta}J_M^-)\sin(2r)+
       J_M^0\cos(2r),
\end{equation}
we obtain
\begin{equation}
    \langle M,\xi,n|N|M,\xi,n\rangle=M\sin^2r+n\cos(2r).
\end{equation}
Similarly we have
\begin{eqnarray}
    \langle M,\xi,n|N^2|M,\xi,n\rangle &=&
    M^2\sin^4r+2Mn\sin^2r\cos(2r)+n^2\cos^2(2r) \nonumber \\
    & & +\frac{1}{4}\sin^2(2r)[(M-n+1)n+(M-n)(n+1)].
\end{eqnarray}
So Mandel's $Q$-index \cite{man} is obtained as
\begin{equation}
    Q=\frac{\langle\Delta N^2\rangle -\langle N\rangle}{
      \langle N \rangle}=
      \frac{-A\sin^4r+(A-M+2n)\sin^2r-n}{(M-n)\sin^2r+n\cos^2r},
\end{equation}
where
\begin{equation}
    A=2Mn-2n^2+M=2n(M-n)+M>0.
\end{equation}
Since the denominator $ (M-n)\sin^2r+n\cos^2r >0 $, we only 
need to consider the numerator 
\begin{equation} 
    Q'\equiv -A\sin^4r+(A-M+2n)\sin^2r-n,
\end{equation}
which is clearly a parabola with respect to $\sin^2r$. The $Q'$ has
a maximum at the point
\begin{equation}
    \sin^2r_{\mbox{\scriptsize max}}=\frac{A-M+2n}{2A}
\end{equation}
and the maximum is
\begin{equation}
    Q'_{\mbox{\scriptsize max}}=\frac{n^2(M-n)^2-n(M-n)}{A}.
\end{equation}

It is obvious that, in the cases $n=0, M$ or in the case $M=2,
n=1$, $Q'_{\mbox{\scriptsize max}}=0$. In those cases, 
the field is sub-Poissonian except at the points
that $Q'$ takes its maximum, namely at the points
$\sin^2r=0, \sin^2r=1$ and $\sin^2r=1/2$ respectively.

In any other cases, $ Q'_{\max}>0 $ and there exist two
$r$ values, say $ r_-$ and $ r_+$, 
\begin{equation}
    \sin^2r_{\pm} = \frac{n(M-n+1) \pm \sqrt{n^2(M-n)^2-n(M-n)}}{
                    2Mn-2n^2+M}, 
\end{equation}
such that $ Q'=0 $. It is easy to see that
$ 0<\sin^2r_- < \sin^2r_+ < 1 $.
We illustrate $ Q' $ as a function of $\sin^2r$ in 
Fig.\,1(su(2) case),
from which we find that
\begin{enumerate}
\item  When $ 0\leq\sin^2r<\sin^2r_- $ or  $ \sin^2r_+<\sin^2r\leq1 $,
       the field is sub-Poissonian.
\item  When  $ \sin^2r_-<\sin^2r<\sin^2r_+ $, the field is 
       super-Poissonian.
\item  When $\sin^2r=\sin^2 r_-$ or $\sin^2r=\sin^2 r_+ $, the
       field is Poissonian.
\end{enumerate}

\subsection{su(1,1) case}

From 
\begin{equation}
D_{11}^{-1}(M,\zeta)K_M^0 D_{11}(M,\zeta)=
      \cosh(2R) K_M^0 +\frac{1}{2}\sinh(2R)\left(
      e^{i\vartheta} K_M^+ + e^{-i\vartheta} K_M^-\right),
\end{equation} 
we can easily obtain the Mandel's $Q$-index
\begin{equation}
    Q=\frac{(4kn+2n^2+2k)\sinh^4R-
    2n(2k+n-1)\sinh^2R-n}{2(k+n)\sinh^2R+n}.
\end{equation}
Since the denominator 
$ 2(k+n)\sinh^2R+n >0 $, we only need to consider the numerator 
\begin{equation} 
    Q'=A\sinh^4R-(A-2k-2n)\sinh^2R-n, \label{519}
\end{equation}
which is also a parabola with respect to $\sinh^2 R$.

The parabola $Q'$ has a minimum $ Q'_{\min} $ 
at the point $\sinh^2 R_{\min}$, where
\begin{eqnarray}
    \sinh^2 R_{\min} &=&-\frac{4kn+2n(n-1)}{2(4kn+2n^2+2k)}
    \geq 0, \nonumber \\
    Q'_{\min} &=& 
    -\left(\frac{(2kn+n^2-n)^2}{4kn+2n^2+2k}+n\right)\leq 0.
\end{eqnarray} 
Since $Q'=-n\leq 0$ for $\sinh^2R=0$, so the parabola has only one
point intersecting with the $\sinh^2R$-axis. We denote this
point by $\sinh^2 R_+$ and it is given by             
\begin{equation}
    \sinh^2 R_+ =
              \frac{-(4kn+2n^2-2n)+\sqrt{(4kn+2n^2-2n)^2 +
              4(4kn+2n^2+2n)n}}{2(4kn+2n^2+2k)}\geq 0.
\end{equation}
We illustrate this parabola in Fig.\,1(su(1,1) case). We first note that in the case 
$n=0$ (negative binomial states),  
$ \sinh^2 R_{\min}=\sinh^2 R_+=Q'_{\min}=0$,
and $ Q'>0 $ which means the field is super-Poissonian except
the point $\sinh^2 R =0$ on which $Q'=0$ and the 
field is Poissonian.

When $n\geq 1$, from Fig.\,1(su(1,1) case), we find that, (1) when 
$0\leq \sinh^2R<\sinh^2R_+$, the field is sub-Poissonian;
(2) when $\sinh^2R=\sinh^2 R_+$, the field is Poissonian;
(3) when $\sinh^2R_+<\sinh^2R $, the field is super-Poissonian.

We find that both su(2) and su(1,1) displaced number states 
exhibit sub-Poissonian ian statistics in some ranges of parameters involved.

\section{Squeezing effect}
\setcounter{equation}{0}
 
Define two quadratures $x$ (coordinate) and $p$ (momentum)
\begin{equation}
       x=(a^\dagger+a)/\sqrt{2},\ \ \ 
       p=(a^\dagger-a)/\sqrt{2}.
\end{equation}
Then their variances are
\begin{eqnarray}
   && (\Delta x^2)\equiv\langle x^2\rangle-\langle x\rangle^2=
      \frac{1}{2}+\langle N \rangle 
      +\mbox{Re}\langle a^2\rangle
      -2\left(\mbox{Re}\langle a\rangle\right)^2, \nonumber \\  
   && (\Delta p^2)\equiv\langle p^2\rangle-\langle p\rangle^2=
      \frac{1}{2}+\langle N \rangle 
      -\mbox{Re}\langle a^2\rangle
      -2\left(\mbox{Im}\langle a\rangle\right)^2.
\end{eqnarray}  
If $(\Delta x)^2 <1/2 $
(or $(\Delta p)^2 <1/2$), we say that the quadrature $x$ 
(or $p$) is squeezed.

For su(2), we have
\begin{eqnarray}
   && \langle a \rangle =e^{i\theta}\sum_{m=0}^{M-1}\sqrt{m+1}
   {\cal D}_{m}(M,n,|\varsigma|){\cal D}_{m+1}(M,n,|\varsigma|), 
   \nonumber\\
   && \mbox{Re}\langle a^2 \rangle =\cos 2\theta\sum_{m=0}^{M-2}
   \sqrt{(m+1)(m+2)}{\cal D}_{m}(M,n,|\varsigma|)
   {\cal D}_{m+2}(M,n,|\varsigma|),\nonumber \\
   && \langle N\rangle = \sum_{m=0}^{M}
   m {\cal D}_{m}(M,n,|\varsigma|)^2,
\end{eqnarray}  
and for su(1,1), 
\begin{eqnarray}
   && \langle a \rangle =e^{i\vartheta}\sum_{m=0}^{\infty}\sqrt{m+1}
   {\cal G}_{m}(M,n,|\lambda|){\cal G}_{m+1}(M,n,|\lambda|), \nonumber\\
   && \mbox{Re}\langle a^2 \rangle =\cos 2\vartheta\sum_{m=0}^{\infty}
   \sqrt{(m+1)(m+2)}{\cal G}_{m}(M,n,|\lambda|)
   {\cal G}_{m+2}(M,n,|\lambda|),\nonumber \\
   && \langle N\rangle = \sum_{m=0}^{\infty}
   m {\cal G}_{m}(M,n,|\lambda|)^2.   
\end{eqnarray}

We first note that $(\Delta x)^2$ and
$(\Delta p)^2$ are related with each other
by the following relation
\begin{equation}
    (\Delta x)^2_{\theta}= (\Delta p)^2_{\theta\pm\pi/2},
    \ \ \ \ \ 
    (\Delta x)^2_{\vartheta}= (\Delta p)^2_{\vartheta\pm\pi/2}.
\end{equation}
So hereafter we only consider the quadrature
$x$. Then, it is easy to see that
$(\Delta x)^2$ is a $\pi$-periodic
function of $\theta$ (or $\vartheta$ for su(1,1) case) 
and it is symmetric with respect to $\theta=\pi/2$
(or $\vartheta=\pi/2$).
 
Fig.\,2 shows how $(\Delta x)^2 $
of the state $ D_2(M,\xi)|n\rangle $ depends on parameters 
$|\varsigma|$ and $\theta$, respectively. From these plots 
we find that
 
1. When $\theta=0$. In the starting point $|\varsigma|=0$
(corresponding to number state $|n\rangle$) 
$(\Delta x)^2=\frac{1}{2}+n$ and the quadrature $x$ is not
squeezed. Then, with the increase of $|\varsigma|$, it becomes 
squeezed drastically until the maximum of squeezing 
(minimum of $(\Delta x)^2$) is reached.  By further increasing 
$|\varsigma|$,  the squeezing becomes weaker and weaker until it
disappears for a large enough $|\varsigma|$. 
The squeezing range depends on $n$: the larger $n$, the 
narrower the squeezing range. For large enough $n$, 
there is no squeezing. 
 
2. Dependence on $\theta$.  Since $(\Delta x)^2$ 
is symmetric with respect to $\theta_m=\pi/2$, so 
we only plot $0\leq\theta\leq \pi/2$ part.  
We see that, with the decrease
of $\theta$ form $0$, the squeezing
becomes weaker and weaker and finally disappears
for large enough $\theta$. 

Fig.\,3 shows how $(\Delta x)^2 $
of the state $ D_{11}(M,\zeta)|n\rangle $ depends on parameters 
$|\lambda|$ and $\vartheta$, respectively. From these plots 
we find that
 
1. When $\vartheta=\pi/2$. In the starting point $|\lambda|=0$
(corresponding to number state $|n\rangle$) 
$(\Delta x)^2=\frac{1}{2}+n$ and the quadrature $x$ is not
squeezed, as in the su(2) case. Then, with the increase of 
$|\lambda|$, it becomes 
squeezed. The larger $|\lambda|$, the stronger the squeezing.
The squeezing range depends on $n$: the larger $n$, the 
narrower the squeezing range. 
 
2. Dependence on $\vartheta$.  We only plot 
$0\leq\vartheta\leq \pi/2$ part.  
We see that, with the decrease
of $\vartheta$ form $\pi/2$, the squeezing
becomes weaker and weaker drastically and 
finally disappears. 

So we find that both su(2) and su(1,1) displaced number 
states exhibit squeezing effect in some ranges of parameters
involved. 
 
\section{Conclusion}
\setcounter{equation}{0}

In this paper we have systematically investigated the
su(2) and su(1,1) DNS and their
various properties. These states are eigenstates of 
some Hamiltonians describing density-dependent 
interaction between single mode radiation field and
classical current. As intermediate states both su(2) 
and su(1,1) DNS degenerate to the 
number and displaced number states in two
different limits. We obtain their explicit expansion
in the Fock space and photon distributions. We
analytically studied their statistical properties and
find that these states are sub-Poissonian states
in some ranges of parameters involved. It is 
of interest that these states exhibit strong squeezing
effects, although
their limiting states, number and displaced number 
states, do not.

\section*{Acknowledgments}

This work is partially supported by the National Natural 
Science Foundation of China through Northeast Normal
University (19875008) and the State Key Laboratory of 
Theoretical and Computational Chemistry, Jilin University.


\end{document}